\input{aipcheck}
\documentclass[
  ]
  {aipproc}
\layoutstyle{6x9}

\begin{document}

\title{Hyper-accreting tori of Gamma Ray Bursters}

\classification{95.30.Qd;  97.10.Gz; 97.60.Lf; 97.60.Bw}
\keywords      {black hole physics -- supernovae: general -- gamma-rays: bursts 
-- methods: numerical -- MHD -- general relativity}

\author{M.V. Barkov}{
  address={Department of Applied Mathematics, The University of Leeds,
Leeds, LS2 9GT, UK}
  ,altaddress={Space Research Institute, 84/32 Profsoyuznaya Street, Moscow
117997, Russia} 
}

\begin{abstract}

We present numerical simulations of axisymmetric magnetised massive
tori around rotating black holes taking into account the energy losses 
due to emission of neutrinos. A realistic equation of state is used  
which takes into account the energy losses due to dissociation of nuclei. 
The heating due to neutrino-antineutrino annihilation is not included. 
We study the cases of optically thick, semi-transparent, and 
optically thin to neutrino discs. 
We show that neutrino cooling does not change significantly the
structure of accretion flow and the total energy release. 
The time scale of accretion is set by the torus angular momentum. 
Due to the lack of magnetic dynamo in our calculations, it is 
the initial strength of magnetic field and its topology that determine 
the process of jet formation and its energetics. 
Extrapolation of our results gives the total energy released in the jet 
$\sim 10^{52}$ erg. This is sufficient to explain the hypernovae explosions 
associated with GRB 980425 and GRB 030329.

\end{abstract}

\maketitle


\section{Introduction}

The long-soft Gamma Ray Bursts (GRB) are found predominantly in the blue light of 
galaxies \citep{bloom02,fruch06,kos05}, that is 
in the region where the most massive stars die. This is supported by the 
debated connection of GRB 980425 with SN 1998bw
\citep{sof98,gal98,pian00} and  the detection of low-redshift
($z=0.1685$; \cite{gre03}) GRB 030329 and its associated supernovae, SN 2003dh
\citep{math03,hjo03,sok03}. Moreover, the afterglow spectra of many GRBs show 
broad SN spectral features.

The most popular model of GRB central engine is based on the ``failed
supernova'' scenario of stellar collapse, or ``collapsar'', where
the iron core of progenitor forms a BH \cite{W93}. 
If the progenitor is non-rotating then its
collapse is likely to continue in a ``silent'' manner until the 
whole star is swallowed by the BH. 
If, however, the specific angular momentum in the
equatorial part of the stellar envelope exceeds that of the last
stable orbit of the BH then the collapse becomes highly anisotropic.
While in the polar region it may proceed more or less uninhibited
the equatorial layers form dense and massive accretion
disk. The gravitational energy released in the disk can be very large,
more then sufficient to stop the collapse of outer layers and drive
GRB outflows, presumably in the polar direction where the mass density is much
lower \citep{MW99,BK07a}. 

The last few years witnessed dramatic progress in general relativistic
magnetohydrodynamic (GRMHD) simulations of BH accretion systems. 
They revealed complex structure that can be decomposed into
a disk, corona, disk wind, and highly magnetised polar region that hosts a 
black hole jet driven by the Blandford-Znajek mechanism 
\citep{BZ77,vhk03,krol06,mac04,mac06}. 
However, all these simulations used very primitive equations of state (EOS) 
and neglected neutrino cooling. The first 
2D GRMHD simulations in Kerr metrics with realistic EOS and neutrino
cooling simulations have been carried out only very recently \cite{shibata} but 
the physical time span of their computations  was rather short, only $\simeq 0.06$~sec.
In this paper we present the results of a similar study whose main aim 
is to investigate the effects of
neutrino cooling processes on the disk structure and dynamics.

\section{Physical processes}
\label{PP}

In these simulations we use realistic EOS  \cite{ard05}, which also accounts for the 
photo-disintegration of nuclei. The cooling rates due to neutrino emission via  
annihilation of $e^+e^-$ pairs, photo-production, and plasma mechanisms 
are taken from \cite{schinder}, via URCA processes from \citep{ivanova}, 
and via synchrotron mechanism from \citep{bezchas}. 

We compare 3 different
cases: 1) without cooling (model DL2A9NC); 2) cooling in optically thin regime
(model DL2A9C); 3) cooling with incorporated opacity effects (model DL2A9MC). 
The first case is not very realistic in the GRB context and we consider it 
only as a reference model. The second case suits well tori with mass of 
few percent of the solar mass. It could be relevant for the models of
neutron star-neutron star and BH-white dwarf mergers. The third case requires
a torus of several solar masses. 

To describe the neutrino opacity we follow \citep{ard05}, namely we multiply 
the local loss rate by the factor $e^{-\tau_{\nu}}$, 
where $\tau_{\nu}=S_{\nu}nl_{\nu}$ is the neutrino optical depth, $n=\rho/m_p$ is the
concentration of baryons, $m_p$ is the proton mass, and  $S_{\nu}$ is the neutrino
interaction cross-section \citep{thom01, tub75}. 
Neglecting the pair loading at high densities 
$S_{\nu}=\left(\left[\kappa_e+\kappa_p\right]Y_e + \kappa_n(1-Y_e)\right) T\epsilon_{\nu}$,
where $\epsilon_{\nu} = 3.15 kT$, $\kappa_e = \sigma_0 \Lambda_e /2m_e^2$, 
$m_e$ is the electron mass in MeV, $\sigma_0\simeq 1.71\times 10^{-44}$ cm$^2$, 
$\Lambda_i=(c_V+c_A)^2+\frac{1}{3}(c_V-c_A)^2$,
where $c_v$ and  $c_A$ are the vector and axial-vector coupling constant for
given neutrino species. $\Lambda_e \simeq 2.2$, 
$\kappa_n=\sigma_0(1+3g^2_A)/(16m_e^2)$ and 
$\kappa_p=\sigma_0[4\sin^4\theta_W - 2\sin^2\theta_W+(1+3g^2_A)/4]/(4m_e^2)$, 
where $\sin^2\theta_W \simeq 0.231$
and $g_A\simeq -1.26$ is the axial-vector coupling constant.
The characteristic length-scale of neutrino absorption $l_{\nu}$ 
is difficult to calculate directly due to the complex and dynamic structure 
of solutions.  Instead we use the following simplification  
$	l_{\nu}=\frac{\rho}{\nabla \rho} \approx \left(0.27+1.5\frac{\rho}{10^{12}}
\right)r_g$, where $r_g = GM_{BH}/c^2$ is the gravitational radius of BH.

At very high densities $\rho > 8.1\times 10^6$ g cm$^{-3}$ the process of
electron capture on nuclei becomes important. In order to take this into account 
we have utilised the results obtained in \citep{bay71b} and made a simple 
approximation which is 7\% accurate.


\section{Simulation setup}
\label{SS}

The simulations are carried out with an upwind
conservative scheme that based on a linear Riemann solver and uses the
constrained transport method to evolve magnetic field.  The details of this
numerical method and various tests are described in \cite{K99,K04b}.
The neutrino cooling is introduced via the source term in the energy-momentum
equation
$$
  \partial_\nu(\sqrt{-g} T^\nu_\mu)=\sqrt{-g}Su_\mu,
$$ 
where $u^\nu$ is the four-velocity of plasma, $g$ is the determinant of 
metric tensor, and $S$ is the cooling rate as measured in the fluid frame. 
The gravitational attraction of BH is introduced via the Kerr metric in
Kerr-Schild coordinates, $\{t,\phi,r,\theta\}$.  The two-dimensional
computational domain is $(r_0<r<r_1)\times(0<\theta<\pi)$, where $r_0 =
(1+\sqrt{1-a^2}/2) \; r_g$, there
$r_g = 14.847$ km and $r_1=200 \;r_g = 2969$ km. The total mass within the
domain is small compared to the mass of BH (less then 25\%) that allows us to
ignore its self-gravity. 
The grid is uniform in $\theta$ where it has 320 cells and almost
uniform in $\log(r)$ where it has 459 cells, the linear cell size being the same
in both directions.

The initial solution describes an equilibrium torus with constant
specific angular momentum $l_0$ \cite{fish76,abr78}. 
The black hole mass $M_{BH} = 10 M_{\odot}$ and its rotation parameter 
$a=0.9$. The torus mass $M_{tor} = 2.55 M_{\odot}$ and its specific 
angular momentum $ l_0 = 2.8 r_g c = 1.25\times 10^{17} \; g \;cm^2 sec^{-1}$. 
This setup corresponds to the collapse of a massive star when its core and 
inner envelope form a black hole and the outer envelope forms an accretion disk. 
The initial magnetic 
field is purely poloidal, thus the only non-zero component of its vector 
potential is $A_{\phi} \propto W(r,\theta)^3$, here $W(r,\theta)=\frac{1}{2}\ln\left|\frac{g_{t\phi}g_{t\phi}-g_{tt}g_{\phi\phi}}{g_{\phi\phi}+2l_0 g_{t\phi}+l_0^2g_{tt}}\right|$. 
The ratio of magnetic to gas pressure $\beta=P_m/P_g$ does not
exceed $3\times 10^{-2}$. 
In the initial solution more then 80\% of the total pressure comes 
from the radiation and only 20\% from the degenerated matter. In these conditions 
the neutrino cooling is important.


\section{Results}
\label{Res}

\begin{figure}
\includegraphics[width=40mm,angle=-90]{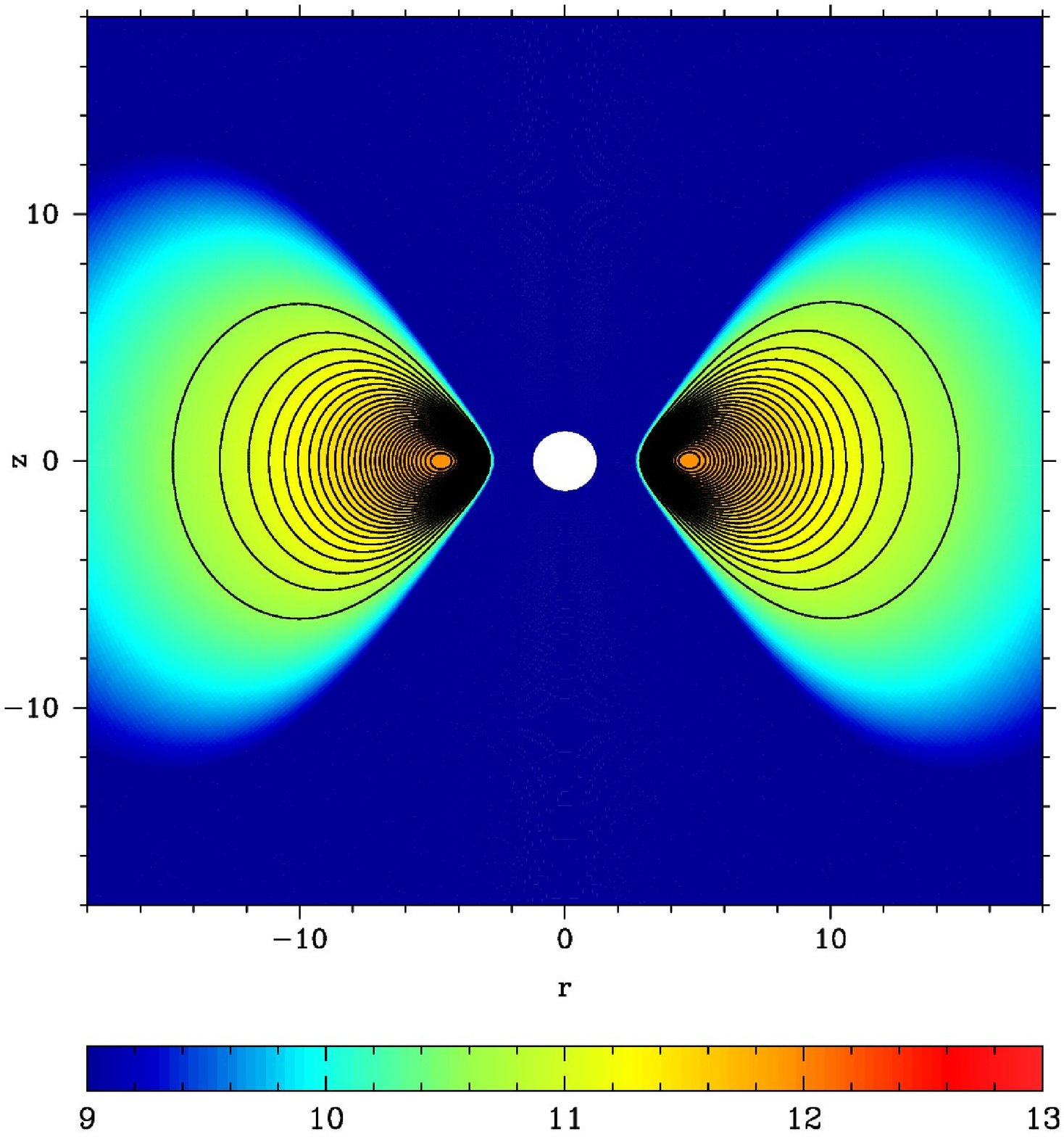}
\includegraphics[width=40mm,angle=-90]{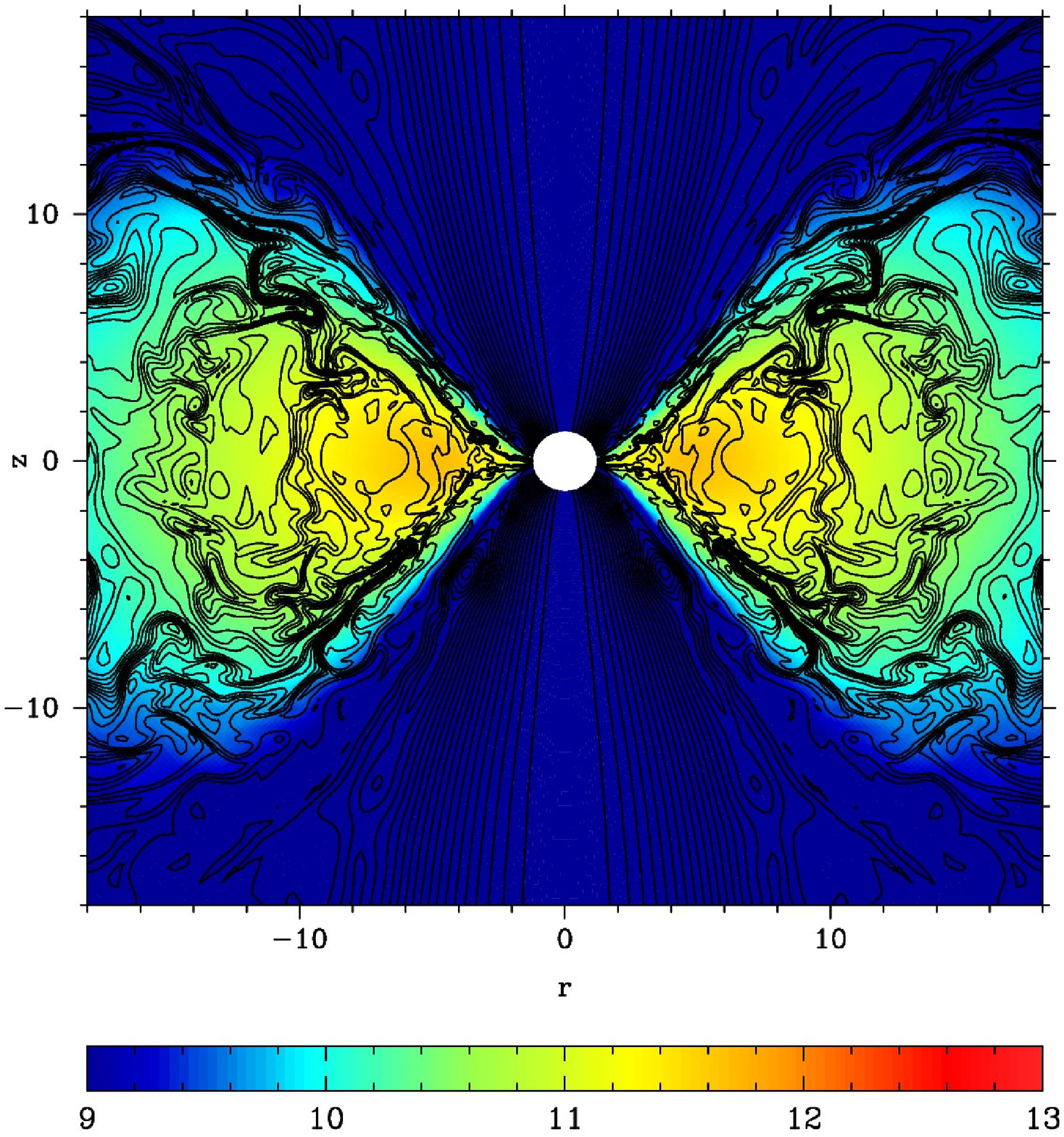}
\includegraphics[width=40mm,angle=-90]{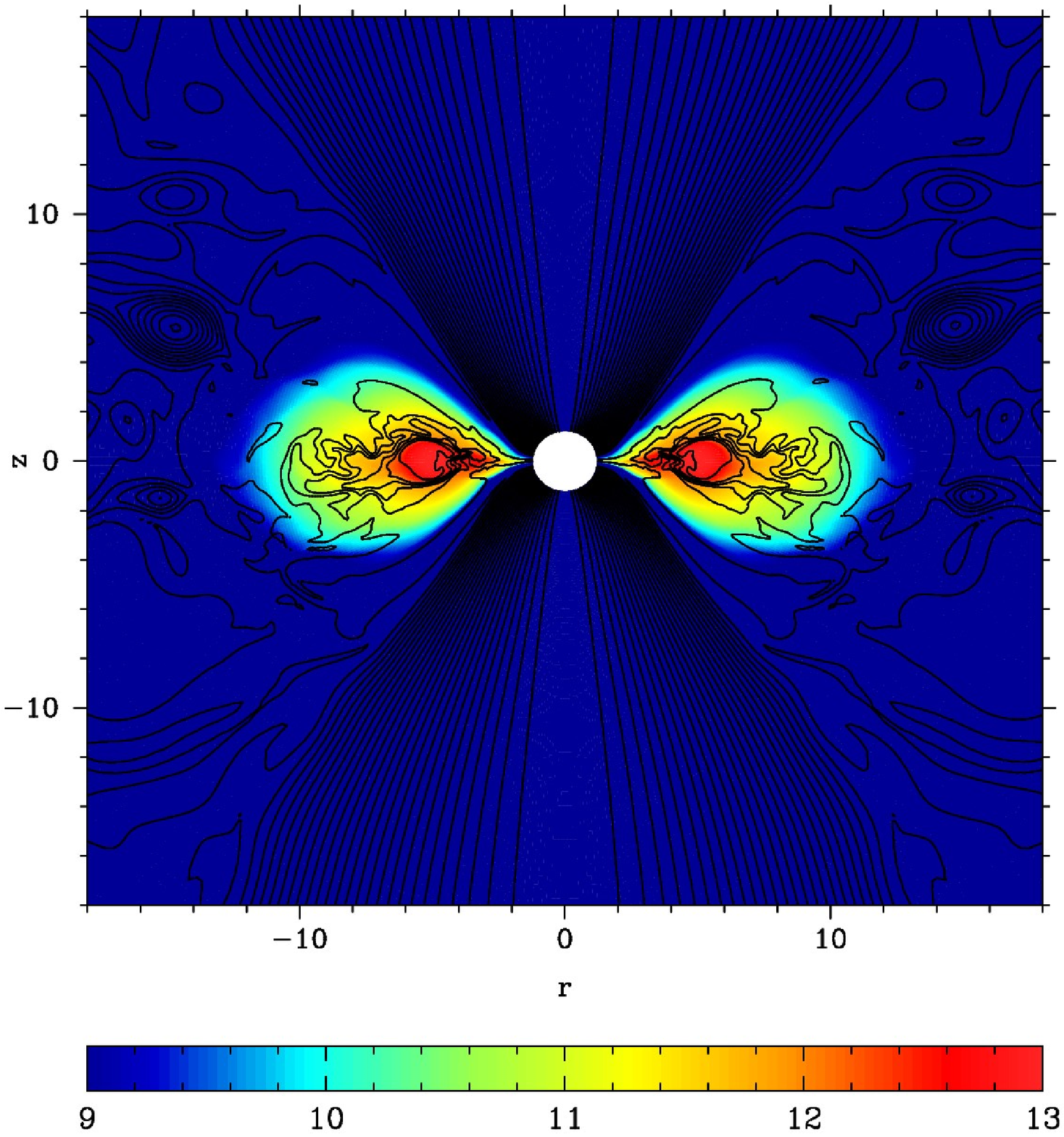}
\includegraphics[width=40mm,angle=-90]{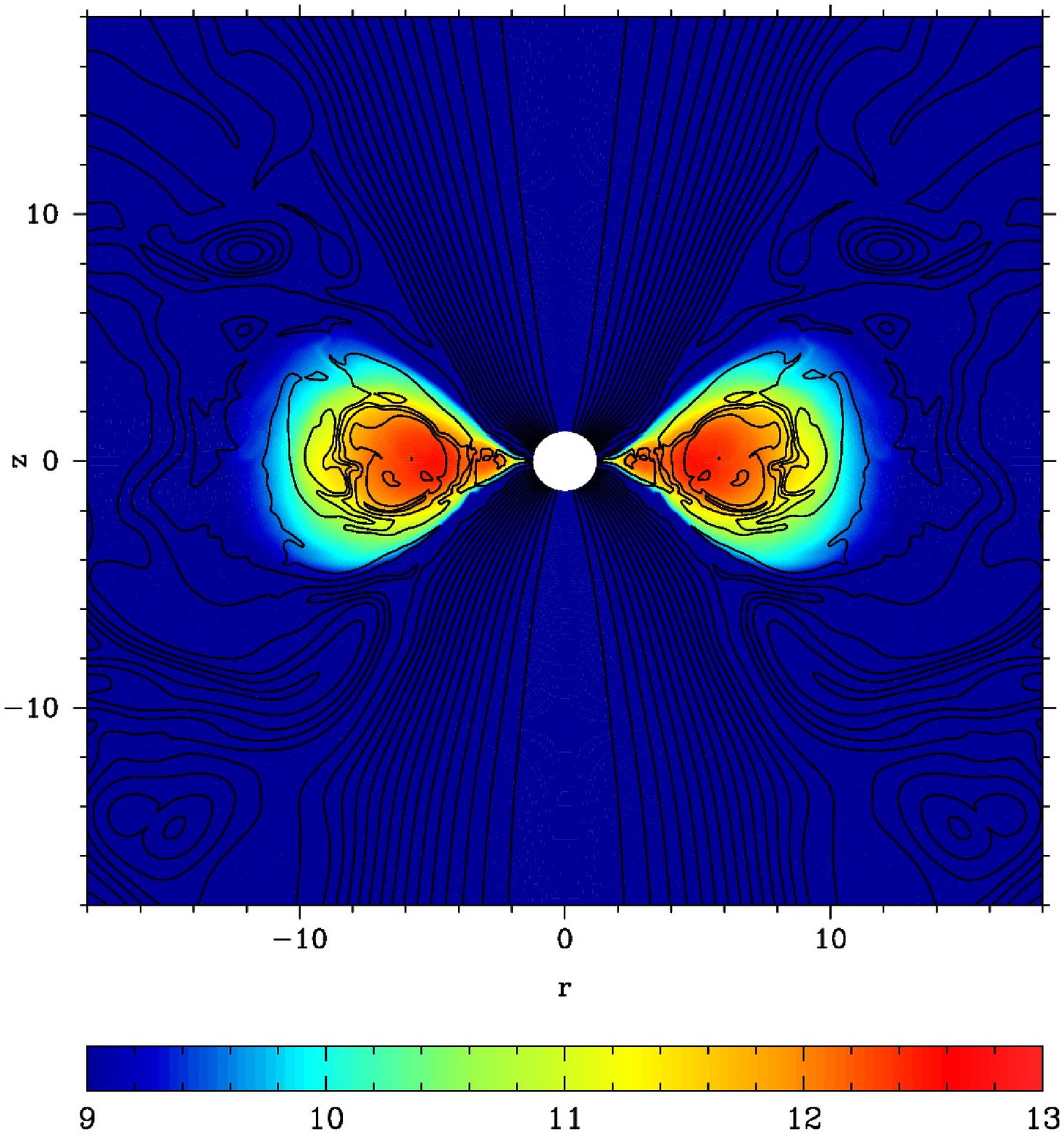}
\caption{$\log_{10}(\rho)$ (colour image) and magnetic field lines.
The initial solution for all models is shown in the upper left panel. 
The other panels from left to right show solutions after 0.2075 sec; model DL2A9NC 
(second on the left), model DL2A9C (second on the right),
model DL2A9MC (first on the right).}
\label{rho_c}
\end{figure}

After initial period of relaxation all our models show the same
structure involving a thick disk, a highly magnetised jet and a wind 
from the disk. 
The fast neutrino cooling in the optically thin model DL2A9C leads to
significant imbalance of the torus. The 
torus collapses to a new state where its thermal pressure becomes
negligible.  In the model DL2A9MC, where we take into account the opacity
effects, the torus interior is optically thick to neutrinos. This
model develops a slightly different structure where the hallo 
is squeezed as in model DL2A9MC but the central core is bigger and has a higher
temperature ($T\approx 2.2\times 10^{11}$ K) compared to the model DL2A9C. 
Figure~\ref{rho_c} shows the impact of neutrino cooling
on the structure of the torus. Without cooling the torus keeps the same
size and shape as in the initial solution. The strong cooling in model 
DL2A9C leads to the collapse of initial configuration with the peak 
density increasing 10-fold and reaching {$10^{13}$ g cm$^{-3}$}. 
The moderate cooling in model DL2A9MC results in lower peak  
density,  {$3\times 10^{12}$ g cm$^{-3}$}.
In DL2A9MC and DL2A9C the accretion flow is thinner than in DL2A9NC (see
fig.\ref{rho_c}). In DL2A9NC the highly magnetised jet region is much smaller 
and more variable than in DL2A9C and DL2A9MC. However, 
the global structure of the accretion flow does not change very much and 
the energy release rate in jets of all models is similar. 
In model DL2A9MC the 
efficiency,
$$
\eta = \int_0^T \dot E_{tot} dt / \int_{0}^T \dot M_{BH} c^2 dt,
$$ 
is about 0.003. In model DL2A9NC it is 3.8 times lower.

\begin{figure}
\includegraphics[width=70mm,angle=0]{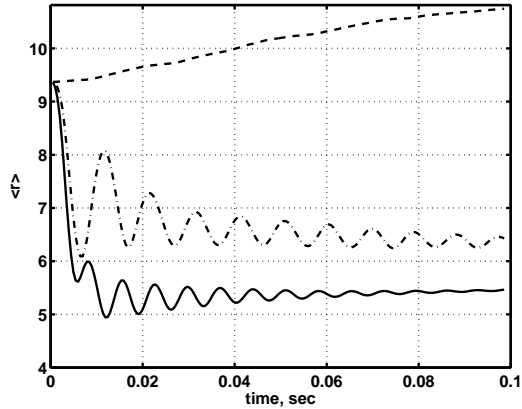}
\caption{Evolution of the mass-averaged torus radius. Dashed line -- DL2A9NC, 
dot-dashed --DL2A9MC, solid -- DL2A9C.}
\label{oscil}
\end{figure}

At the start of simulations the central regions of models with cooling 
loose pressure support and begin to fall into the potential well. This 
leads to the oscillations similar to those reported 
in \citep{zan05}. However, we observe fast dumping of these oscillations 
(see fig. \ref{oscil}) One reason is the strong radiative shocks driven into 
the outer layers of the torus which are slow to react to the loss of 
equilibrium. The other reason is Maxwell stresses. 
In the optically thin case (model DL2A9C) the initial amplitude of oscillations 
${dr^m/r^m}=0.05$, where $r^m = \int_M rdm / \int_M dm$ is the 
mass-averaged torus radius. The oscillations degrade quickly with the quality
factor  $Q=|(dr_i^m+dr^m_{i+1})/2(dr^m_i-dr^m_{i+1})|=3.8$, here $dr^m_i$ is maximal
deviation from equilibrium of i-th oscillation. 
The estimated relaxation time
is $t_{relax}=t_{osc}\times Q \approx 0.027$ sec. In model DL2A9MC with 
less effective cooling the initial amplitude of oscillations is higher, 
${dr^m/r^m}=0.15$, but the quality factor is lower, $Q=1.6$. 
After few oscillations the amplitude decreases to ${dr^m/r^m}=0.05$, 
the effective viscosity decreases and the quality factor grows to $Q = 6.7$.  
The corresponding relaxation time $t_{relax}\approx 0.063$ sec. 
When neutrino cooling is not included the oscillations do not appear at all 
and the mass-averaged radius of the torus increases in time due to the 
development of wind.

\begin{table}
\begin{tabular}{l c c c c c c c}
\hline
 model  & $l$ & $a$ & $\dot M_{BH}$  & $\dot M_{w} $   & $E^{tot}_{51}$ &
$\eta$ &\\
\hline
 DL2A9NC   & 2.8 & 0.9 & 1.3937 & 0.0189   & 1.8530 & 0.00074\\
 DL2A9C    & 2.8 & 0.9 & 0.8588 & 0.0134   & 2.9039 & 0.0019 \\
 DL2A9MC   & 2.8 & 0.9 & 0.512  & 0.0276    & 2.577  & 0.0028 \\
\hline
\label{tab2}
\end{tabular}
\caption{Main results.}
\end{table}


\section{Discussion and conclusions}
\label{Dis}

In all our models we observe the development of a Poynting-dominated jet from BH.
The neutrino cooling processes do not have a significant effect on the jet 
power. In fact it appears to be less disrupted due to interactions 
with disk wind and corona. On the other hand, the disk wind weakens when the 
cooling is included. 
The estimated total energy release during the whole period of accretion 
$E_{tot} \approx \eta M_{torus}c^2 \sim
10^{52}$ erg. This can explain the energetics of hypernovae explosions associated 
with GRB 980425 and GRB 030329. The studies of magnetic acceleration of 
Poynting-dominated jets show that high terminal Lorentz factors can be achieved 
together with effective conversion of electromagnetic energy into the bulk 
motion energy of the jet \cite{KBVK07,BK08}. Thus, this is a promising model 
of central engines of Gamma Ray Bursts.     
 
Our model is complimentary to the model of magnetar driven GRBs and 
hypernovae, e.g. \cite{KB07}. Both these models explain the origin of collimated 
GRB jets and power of hypernovae. Further investigation is required to 
distinguish these two scenario and determine their specific observational 
signatures. At the moment we can only say that the chemical composition of the 
jet will different. In the collapsar case we expect highly Poynting dominated
 $e^+e^-$ jets from onset whereas in the magnetar model the jets are initially 
strongly contaminated by baryons. 

The disc oscillations observed in our simulations are the product 
of initial setup because the effects of neutrino cooling have not been 
included in the initial equilibrium solution. They are quickly dumped and 
thus unlikely to be excited in realistic astrophysical conditions. This 
conclusion is supported by the simulations where the formation of accretion 
disks is followed from the onset of stellar collapse \cite{MW99,BK07a}. 
     
\begin{theacknowledgments}
The author would like to thank S.S. Komissarov and G.S.Bisnovatyi-Kogan for help
in calculations and discussions. This calculations were carried out on St
Andrews UK MHD cluster and the White Rose Grid facilities. This research is
funded by PPARC/STFC under the rolling grant ``Theoretical Astrophysics in
Leeds''.
\end{theacknowledgments}



\bibliographystyle{aipproc}   


\begin{thebibliography}{99}

\bibitem{bloom02} Bloom J.S., Kulkarni S.R., Djorgovski S.G., \emph{Astronomical
Journal} \textbf{123}, 1111--1148 (2002)

\bibitem{fruch06} Fruchter A.S., Levan A.J., Strogler L.,
Vreeswijk P.M., et~al., \emph{Nature} \textbf{441}, 7092, 463--468 (2006)

\bibitem{kos05} Blinnikov S.I., Postnov K.A., Kosenko D.I.,
Bartunov O.S., \emph{Astron. Lett.} \textbf{31}, 6, 365--374 (2005)

\bibitem{sof98}	Soffitta P., Feroci M., Piro L., Zand J., Heise J. et~al.,
\emph{IAU Circ} \textbf{6884}, 1 (1998)

\bibitem{gal98} Galama T.J., Vreeswijk P.M., van Paradijs
J. et~al., \emph{Nature} \textbf{395}, 6703, 670--672 (1998)

\bibitem{pian00} Pian E., Amati L., Antonelli L.A., Butler
R.C. et~al., \emph{Astrophysical Journal} \textbf{536}, 778--787 (2000)

\bibitem{gre03} Greiner J., Peimbert M., Estaban C. et~al., \emph{GCN}
\textbf{2020},  (2003)

\bibitem{math03} Matheson T., Garnavich P.M., Stanek
K.Z. et~al., \emph{Astrophysical Journal} \textbf{599}, 394--407 (2003)

\bibitem{hjo03} Hjorth J., Sollerman J., M\o{}ller P., Fynbo
J.P.U. et~al., \emph{Nature} \textbf{423},  6942, 847--850 (2003)

\bibitem{sok03} Sokolov V.V., Fatkhullin T.A., Komarova
V.N. et~al., \emph{Bull. Spec. Astroph. Obser.} \textbf{56}, 5--14 (2003)

\bibitem{W93} Woosley S.E., \emph{Astrophysical Journal} \textbf{405}, 273--277
(1993)

\bibitem{MW99} MacFadyen A.I. \& Woosley S.E., \emph{Astrophysical Journal}
\textbf{524}, 262--289 (1999)

\bibitem{BK07a} Barkov M.V., Komissarov S.S., \emph{Month. Not. Royal
Astron. Soc.: Lett.} \textbf{385}, L28--L32 (2008)

\bibitem{vhk03} De Villiers J.P., Hawley J.F., Krolik J.H., \emph{Astrophysical
Journal} \textbf{599}, 1238--1253 (2003)

\bibitem{krol06} Hawley J.F., Krolik J.H., \emph{Astrophysical
Journal} \textbf{641}, 103--116 (2006)

\bibitem{mac04} McKinney J.C., Gammie C.F., \emph{Astrophysical
Journal} \textbf{611}, 977--995 (2004)

\bibitem{mac06} McKinney J.C., \emph{Month. Not. Royal Astron. Soc.}
\textbf{368}, 1561-1582 (2006)

\bibitem{BZ77} Blandford R.D. \& Znajek R.L., \emph{Month. Not. Royal Astron.
Soc.} \textbf{179}, 433--456 (1977)

\bibitem{shibata} Shibata M., Sekiguchi Yu, Takahashi R., \emph{Progress of
Theor. Physics} \textbf{118}, 2, 257--302 (2007)

\bibitem{ard05} Ardeljan N.V., Bisnovatyi-Kogan G.S.,et~al., \emph{Month.
Not. Royal Astron. Soc.} \textbf{359}, 333--344 (2005)

\bibitem{schinder} Schinder P. J., Schramm D. N., Wiita P. J., et~al.,
\emph{Astrophysical Journal} \textbf{313}, 531--542 (1987)

\bibitem{ivanova} Ivanova L.N., Imshennik V.S., Nadezhin D.K., \emph{Nauchnye
Informatsii} \textbf{13}, 3 (1969)

\bibitem{bezchas} Bezchastnov V.G., Haensel P., et~al., \emph{Astronomy and
Astrophysics} \textbf{328}, 409--418 (1997)

\bibitem{thom01} Thompson T.A., Burrows A., Meyer B.S., \emph{Astrophysical
Journal} \textbf{562}, 887--908 (2001)

\bibitem{tub75} Tubbs D.L., Schramm D.N., \emph{Astrophysical Journal}
\textbf{201}, 467--488 (1975)

\bibitem{bay71b} Baym G., Pethick Ch., Sutherland P., \emph{Astrophysical
Journal} \textbf{170}, 299--306 (1971)

\bibitem{nag07} Nagataki S., Takahashi R., Mizuta A., Takiwaki
T., \emph{Astrophysical Journal} \textbf{659}, 512--529 (2007)

\bibitem{bir07} Birkl R., Aloy M.A., Janka H.Th., M\"{u}ller E., \emph{Astronomy
and Astrophysics} \textbf{463}, 51--67 (2007)

\bibitem{K99} Komissarov S.S., \emph{Month. Not. Royal Astron. Soc.}
\textbf{303}, 343--366 (1999)

\bibitem{K04b} Komissarov S.S., \emph{Month. Not. Royal Astron. Soc.}
\textbf{350}, 1431--1436 (2004)

\bibitem{fish76} Fishbone L.G., Moncrief V., \emph{Astrophysical Journal}
\textbf{207}, 962--976 (1976)

\bibitem{abr78} Abramowicz M., Jaroszynski M., Sikora M., \emph{Astronomy and
Astrophysics} \textbf{63}, 221--224 (1978)

\bibitem{zan05} Zanotti O., Font J.A., Rezzolla L., Montero P.J., \emph{Month.
Not. Royal Astron. Soc.} \textbf{356}, 1371--1382 (2005)

\bibitem{KBVK07} Komissarov S.S., Barkov M.V., Vlahakis N., K\"onigl A.,
\emph{Month. Not. Royal Astron. Soc.} \textbf{380}, 51--70 (2007)

\bibitem{BK08} Barkov M.V., Komissarov S.S.,  
\emph{preprint: astro-ph} \textbf{0801.4861}, 1--4 (2008)

\bibitem{KB07} Komissarov S.S., Barkov M.V., \emph{Month. Not. Royal Astron.
Soc.} \textbf{382}, 1029--1040 (2007)



\end{thebibliography}

\end{document}